# Understanding LiOH Chemistry in a Ruthenium Catalyzed Li-O$_2$ Battery

Tao Liu,[a] Zigeng Liu,[a] Gunwoo Kim,[a] James T. Frith,[b] Nuria Garcia-Araez,[b] Clare P. Grey*[a]

**Abstract:**

Non-aqueous Li-O$_2$ batteries are promising for next generation energy storage. New battery chemistries based on LiOH, rather than Li$_2$O$_2$, have recently been reported in systems with added water, one using a soluble additive LiI, and the other using solid Ru catalysts. Here, we focus on the mechanism of Ru-catalyzed LiOH chemistry. Using nuclear magnetic resonance, operando electrochemical pressure measurements and mass spectrometry, we show that on discharging LiOH forms via a 4 e⁻ oxygen reduction reaction, the H in LiOH coming solely from added H$_2$O and the O from both O$_2$ and H$_2$O. On charging, quantitative LiOH oxidation occurs at 3.1 V, with O being trapped in a form of dimethyl sulfone in the electrolyte. Compared to Li$_2$O$_2$, LiOH formation over Ru incurs hardly any side reactions, a critical advantage for developing a long-lived battery. An optimized metal catalyst-electrolyte couple needs to be sought that aids LiOH oxidation and is able stable towards attack by hydroxyl radicals.

Non-aqueous Li-O$_2$ batteries possess a high theoretical energy density, 10 times higher than that of the current lithium ion batteries.[1] There have been considerable efforts from academia and industry in the past decade to understand and realize the battery system. Despite of the much research investment, significant challenges remain. One of the most fundamental problems concerns the side reactions that occur during cell cycling.[2] During battery discharge, O$_2$ is reduced to form Li$_2$O$_2$ via an intermediate LiO$_2$;[3] on charging Li$_2$O$_2$ decomposes releasing O$_2$.[4] Both the superoxide and peroxide (either as solvated ions or solid phases) are highly reactive and their formation/decomposition can cause electrolyte and electrode decomposition,[5] especially in the presence of high overpotentials. As a result, many groups have been searching for new Li-O$_2$ battery chemistries.[6-8]

Recently, has been identified as the major discharge product in a couple of Li-O$_2$ battery systems and reversible electrochemical performance has been shown.[7,8] One case is published by some of the authors,[7] concerns the use of a soluble catalyst LiI, which catalyzes the LiOH formation with its H source solely coming from added H$_2$O in the electrolyte; a subsequent study[9] confirmed the proposed 4 e⁻ oxygen reduction reaction (ORR) on discharging. It was also shown on charging the LiOH can be removed with the aid of LiI$_3$ at around 3.1 V.[7] The other case employs a Ru-based solid catalyst in a water-added dimethyl sulfoxide (DMSO) or tetraglyme electrolyte.[8] Ru was proposed to catalyze LiOH formation and decomposition in a tetraglyme electrolyte with 4600 ppm water. In the DMSO case, it was suggested that at low water contents (~150 ppm), a mixture of Li$_2$O$_2$ and LiOH was formed on discharge, and that on charging, Li$_2$O$_2$ is first converted to LiOH, the latter then getting decomposed by Ru catalysts at voltages of as low as ~3.2 V. At higher water contents (~250 ppm), LiOH formation appeared to be dominant on discharge.[10] It is clear that understanding the formation and decomposition of LiOH is not only critical in helping realize a LiOH-based Li-O$_2$ battery, but fundamental insight into LiOH based chemistries may also aid in the development of Li$_2$O$_2$-based batteries that operate utilizing air (or moist oxygen), where LiOH inevitably forms.

In this article, we develop a mechanistic understanding of the Ru-catalyzed oxygen chemistry. Using quantitative nuclear magnetic resonance and operando electrochemical pressure and mass spectrometry measurements, we show that on discharging, a total of 4 electrons per O$_2$ is involved in LiOH formation, this process incurring fewer side reactions compared to Li$_2$O$_2$. On charging, the LiOH is quantitatively removed at 3.1 V, with the oxygen being trapped in the form of soluble dimethyl sulfone in the electrolyte.

The preparation of the Ru/Super P (SP) carbon electrode is described in the Supplementary Materials. Microscopy and diffraction experiments show that Ru crystals of less than 5 nm are well dispersed on the SP carbon substrate (S1). Fig. 1A shows typical electrochemical profiles of Li-O$_2$ batteries prepared using Ru/SP electrodes with various concentrations of added water in a 1 M LiTFSI/DMSO (lithium bis (trifluoromethane) sulfonimide in dimethyl sulfoxide) electrolyte. In the nominally anhydrous case, discharge and charge plateaus are observed at 2.5 and 3.5 V respectively, where an electrochemical process involving two-electrons per oxygen molecule and Li$_2$O$_2$ formation dominates process on discharging (S2). As the water content increases, it is clear (Fig. 1A) that the voltage gaps between discharge and charge reduce considerably. With 50,000 ppm water, the cell discharged at 2.85 V charges at 3.1 V, although further increasing the water content then widens the voltage gaps (S3). Fig. 1B shows the electrochemistry of cells made using various metal catalysts and 1 M LiTFSI/DMSO electrolyte with 4000 ppm water. Although the discharge voltages are all similar, and close to 2.7 V, clear differences are observed on charging, where Ir, Pd, Pt all show charging voltages beyond 3.5 V while for Ru it is only 3.2 V, demonstrating the crucial role of metal catalysis on the charging process. Examining the discharged Ru/SP electrodes, two distinct morphologies were observed for the discharge product (Fig. 1C,D): at lower water contents (e.g. 4000 ppm), cone-shaped particles dominate whereas at higher water contents (e.g. 50,000 ppm), flower-like large agglomerates formed; these morphologies were observed before for LiOH crystals.[7] Indeed, both x-ray diffraction (XRD) and Raman measurements suggest that in the current Ru-based system, LiOH is the only discharge product observed with 4000-50,000 ppm added water; no evidence of other chemical species commonly observed in Li-$O_2$ batteries, such as $Li_2O_2$, $Li_2CO_3$, and $HCOOLi$, is seen by XRD and Raman spectroscopy (Fig. 1E,F). Ir, Pd catalysts also invariably lead to LiOH formation (S4).

[a]    Dr. T. Liu, Dr. Z. Liu, Dr. G. Kim, Prof. C.P. Grey
       Department of Chemistry,
       University of Cambridge
       Lensfield Road, Cambridge, UK CB2 1EW
       *E-mail: cpg27@cam.ac.uk
[b]    Dr. J.T. Frith, Dr. N. Garcia-Araez
       Department of Chemistry,
       University of Southampton
       Highfield Campus, Southampton, UK SO17 1BJ

To demonstrate that at water levels beyond 4000 ppm, LiOH is formed from $O_2$ reduction rather than from electrolyte decomposition, we performed NMR experiments with isotopically labeled H ($D_2O$) and O ($H_2^{17}O$, $^{17}O_2$) (Fig. 2A-C). When natural abundance DMSO and $H_2O$ were used, a dominant $^1H$ NMR resonance at -1.5 ppm attributed to LiOH was observed (Fig. 2A).[7,11] Using $D_2O$, we found a distinct 1st-order quadrupolar-broadened line shape for LiOD in the $^2H$ NMR spectrum (Fig. 2B);[7] when deuterated $d_6$-DMSO and $H_2O$ were used, hardly any LiOD signal was seen (Fig. 2B) and LiOH was the prevailing product (Fig. 2A). The proton in LiOH thus comes overwhelmingly from the added water in the DMSO electrolyte. Next, we $^{17}O$ enriched either gaseous $O_2$ or $H_2O$ to verify the O source in LiOH. In both cases, the resulting $^{17}O$ NMR spectra (Fig. 2C) revealed a resonance at around -50 ppm with a characteristic 2nd-order quadrupolar line shape, which is ascribed to LiOH.[11] It is thus clear that both oxygen atoms in $O_2$ and $H_2O$ contribute to the formation of LiOH, consistent with a 4 electron ORR.

To further verify this mechanism, *operando* pressure measurements show that the recorded pressure matches well with the trend line expected for 4 $e^-$ per $O_2$. Therefore, we propose an overall discharge reaction as follows: (1) $O_2 + 4e^- + 4Li^+ + 2H_2O \rightarrow 4LiOH$. Up to 4 electrons can be stored per $O_2$ molecule, the theoretical capacity of the battery operating via reaction (1) being 1117 mAh/$g_{LiOH}$, comparable to $Li_2O_2$ (1168 mAh/$g_{Li2O2}$). To examine the role of Ru in LiOH formation further, we discharged a SP electrode in a 1 M LiTFSI/DMSO electrolyte with 4000 ppm water. XRD and SEM show that the discharge leads to mainly $Li_2O_2$ formation with an $e^-/O_2$ ratio of 2.2 (S5), whereas discharging Ru/SP in the same electrolyte forms only LiOH. This contrasting behavior suggests that in the absence of Ru the reaction between $H_2O$ and $Li_2O_2$, (2) $2Li_2O_2 + 2H_2O \rightarrow 4LiOH + O_2$, is slow, even though it is thermodynamically favorable ($\Delta G° = -149.3$ kJ/mol) but Ru clearly promoted the LiOH formation. By exposing a Ru/SP electrode discharged in a nominally dry electrolyte (where $Li_2O_2$ is the main product) to the 4000 ppm water-added electrolyte, XRD (S5) shows that all the $Li_2O_2$ was converted to LiOH in the presence of Ru after 10 hours (same time period as used for the galvanostatic discharge in the SP cell); this indicates that Ru can catalyze the reaction (2) above. It is likely that the electrochemical formation of LiOH in the Ru/SP system proceeds via first $Li_2O_2$ generation ($O_2 + 2e^- + 2Li^+ \rightarrow Li_2O_2$) and then Ru catalyzes the chemical reaction of $Li_2O_2$ with $H_2O$ to eventually form LiOH (Reaction 2); overall the reaction is $O_2 + 4e^- + 4Li^+ + 2H_2O \rightarrow 4LiOH$.

Importantly, the LiOH formation during discharge involves few parasitic reactions. Quantitative $^1H$ solid-state NMR spectra (Fig. 2E) comparing the discharged electrodes generated from an anhydrous electrolyte versus those with 4000 and 50,000 ppm added water shows that the $Li_2O_2$ chemistry (at the anhydrous conditions) clearly generated Li formate, acetate, methoxide side reaction products (signified by 0-10 ppm resonances),[7,11] whereas only a single resonance at -1.5 ppm was seen in the LiOH chemistry; similar results were observed with the other metal catalysts (S3). In addition, we found that soaking LiOH in dimethoxyethane (DME) and DMSO for a month showed no change in its solid state NMR spectra (Fig. 2F), indicating that LiOH is chemically inert in these solvents. $^1H$ and $^{13}C$ solution NMR measurements of the electrolytes after discharging and soaking with LiOH under $O_2$ also show that hardly any soluble side-reaction product is detected in the electrochemical LiOH formation (S6).

Now moving to battery charging, this process was characterized by *ex-situ* NMR and XRD measurements of electrodes after multiple cycles, as presented in Fig. 3A-C. They all consistently show that quantitative LiOH formation on discharging and LiOH removal (*even at 3.1 V*) on charging are the prevailing processes during cell cycling. Hardly any residual solid, side-reaction products accumulate in the electrode over extended cycles. Typically, the cells can cycle over 100 cycles at 1 mAh/$cm^2$ (0.5 mAh or 1250 mAh/$g_{Ru+C}$ per cycle), with very consistent electrochemical profiles (S7). Although the *ex-situ* tests supported a highly reversible $O_2$ electrochemistry, *operando* electrochemical pressure and mass spectrometry experiments suggested otherwise: very little gas was evolved on charging (Fig. 3D,E) and the pressure of cell continues to drop over extended cycles (Fig. 2F); these observations imply that oxygen must be trapped and accumulated after charging in the cell, likely in the electrolyte.

Further solution NMR measurements were performed on electrolyte samples prepared from several cycled cells extracted following different cycle numbers, where $^{17}O$ enriched $H_2O$ ($H_2^{17}O$) was used in the electrolyte. Fig. 4 shows the $^1H$ (A), $^{13}C$ (B) and $^{17}O$ (C), and $^1H$-$^{13}C$ heteronuclear single quantum correlation (D) solution NMR spectra of the cycled electrolytes. A common feature is that new peaks at 2.99 ppm ($^1H$), 42.6 ppm ($^{13}C$) and 169 ppm ($^{17}O$) appeared and progressively intensified compared to with cycle number; these resonances consistently point towards the formation of dimethyl sulfone ($DMSO_2$), its identity being further corroborated in the $^1H$-$^{13}C$ correlation spectrum. Of note, the $^{17}O$ signal of $DMSO_2$ is even stronger than the large amount of natural abundance (NA) DMSO used in the solution NMR experiment, suggesting that $DMSO_2$ is likely to be $^{17}O$-enriched. Its growth in intensity is accompanied by the decrease of $H_2^{17}O$, indicating that some $^{17}O$ from $H_2^{17}O$ ended up in $DMSO_2$ due to isotope scrambling in the charging process. Given that LiOH is quantitatively formed and then removed on charge (Fig.3), we propose that the charging reaction is initiated by electrochemical LiOH oxidation to produce hydroxyl radicals, which then *chemically* react with DMSO to form $DMSO_2$: (3) $LiOH \rightarrow Li^+ + e^- + \,^\cdot OH$ (hydroxyl radicals); (4) $DMSO + 2\,^\cdot OH \rightarrow DMSO_2 + H_2O$. The overall reaction thus is: (5) $2DMSO + 4LiOH \rightarrow 2DMSO_2 + 2H_2O + 4e^- + 4Li^+$. It is seen that the same number of electrons is involved in discharge (reaction 1) and charge (reaction 5), with one O reacting per two electrons (as expected for $O_2$ evolution reaction, OER). The electrochemical process, reaction (3), sets the voltage observed on charge, rather than the overall reaction (5). It is known that the formation of surface adsorbed hydroxyl species is the first reaction intermediate on many OER metal catalysts in aqueous media.[12] The added water in the current electrolyte could promote LiOH dissolution, and thus facilitate the access of Ru surfaces to soluble LiOH species resulting in the formation of surface hydroxyl species. Once the radical is formed on charging, it is consumed by reacting with DMSO to form $DMSO_2$ and thus the battery can be continuously charged at a low voltage until all solid LiOH products are removed (see further discussion in S8). The resulting $DMSO_2$ is soluble in the DMSO electrolyte and will not immediately impede ion diffusion or interfacial electron transfer as other insoluble by-products would do, which is perhaps why this side reaction does not rapidly lead to battery failure.

In summary, we have shown that with added water (beyond 4000 ppm) in the electrolyte, the Ru-catalyzed battery chemistry changes from $Li_2O_2$ to LiOH formation, similar reactions being seen for several other metal catalysts. The cell discharge reaction consumes four-electron per reduced $O_2$ molecule. This LiOH formation process involves very few side reactions and LiOH itself is much more stable in organic solvents than $Li_2O_2$; these are the fundamental prerequisite for a long-lived Li-$O_2$ battery. On charging, the Ru quantitatively catalyzes LiOH removal via $DMSO_2$ formation rather than $O_2$ evolution. We propose that $DMSO_2$ forms by the reaction of hydroxyl radicals with DMSO, the former being generated on Ru catalyst surfaces. This work highlights the advantage of using metal catalysts to catalyze a 4 e⁻ ORR with very few side reactions, and also the unique role of a metal catalyst in promoting LiOH formation versus electrolyte decomposition. An optimized catalyst-electrolyte couple needs to be sought for to satisfy both activity towards LiOH oxidation and stability against electrolyte decomposition on charge. This work provides a series of key mechanistic insights into the Ru-catalyzed Li-$O_2$ battery in the presence of water, which will aid the design of catalyst and electrolyte systems that can be used in more practical batteries.

## Experimental Section

Experimental Details: see Supplementary Materials.

## Acknowledgements

The authors thank EPSRC-EP/M009521/1 (TL, GK, CPG), Innovate UK (TL), Darwin Schlumberger Fellowship (TL) and EU Horizon 2020 GrapheneCore1-No.696656 (GK, CPG) for research funding. NGA and JTF thank EPSRC (EP/N024303/1, EP/L019469/1), Royal Society (RG130523) and the European Commission (FP7-MC-CIG Funlab, 630162) for research funding.

**Keywords:** Li-$O_2$ batteries • oxygen reduction/evolution reaction • LiOH • dimethyl sulfone • ruthenium catalysis

**Entry for the Table of Contents** (Please choose one layout)

Layout 1:

## COMMUNICATION

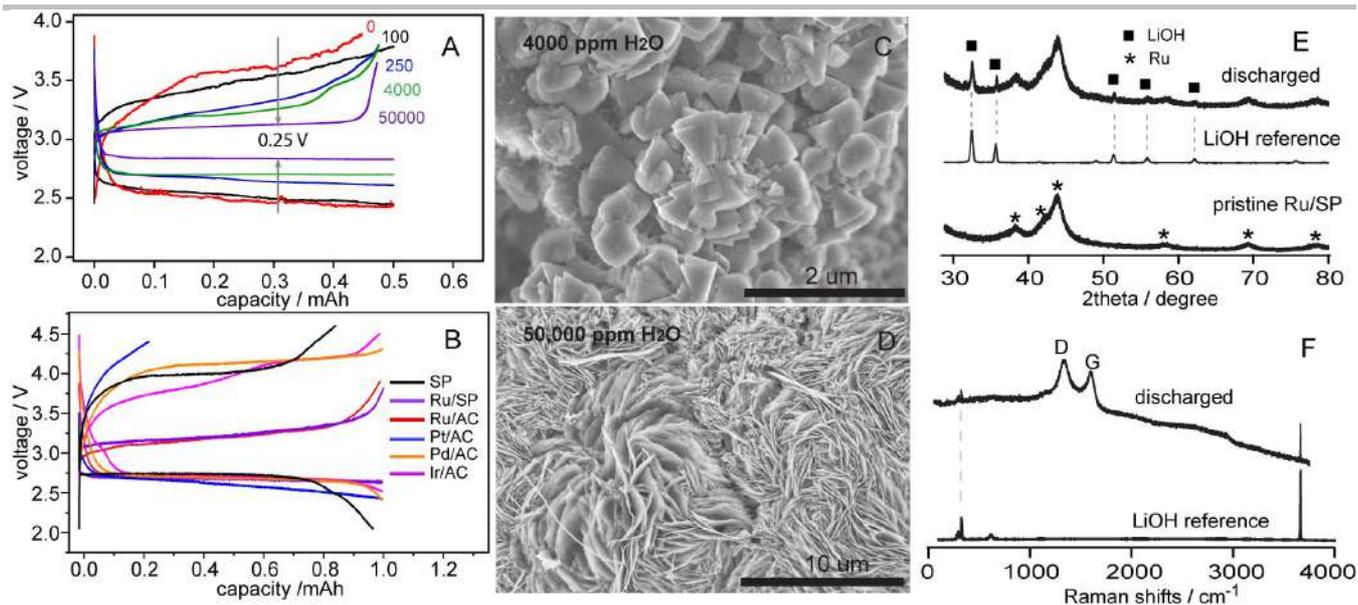

Fig. 1 Electrochemical profiles of Li-O₂ cells with different water contents (in ppm) (A) in a 1M LiTFSI/DMSO electrolyte and using different metal catalysts (AC = activated carbon, SP = super P) (B). Characterization of discharged electrodes by SEM (C,D), XRD (E) and Raman spectroscopy (F). All cells in A use Ru/SP electrodes; All cells in B contain a water content in the electrolyte of 4000 ppm. All cells were cycled at a current of 50 µA (0.1 mA/cm²). The discharged electrodes measured in XRD (E) and Raman (F) are both prepared using the electrolyte with 4000 ppm water and Ru/SP electrodes.

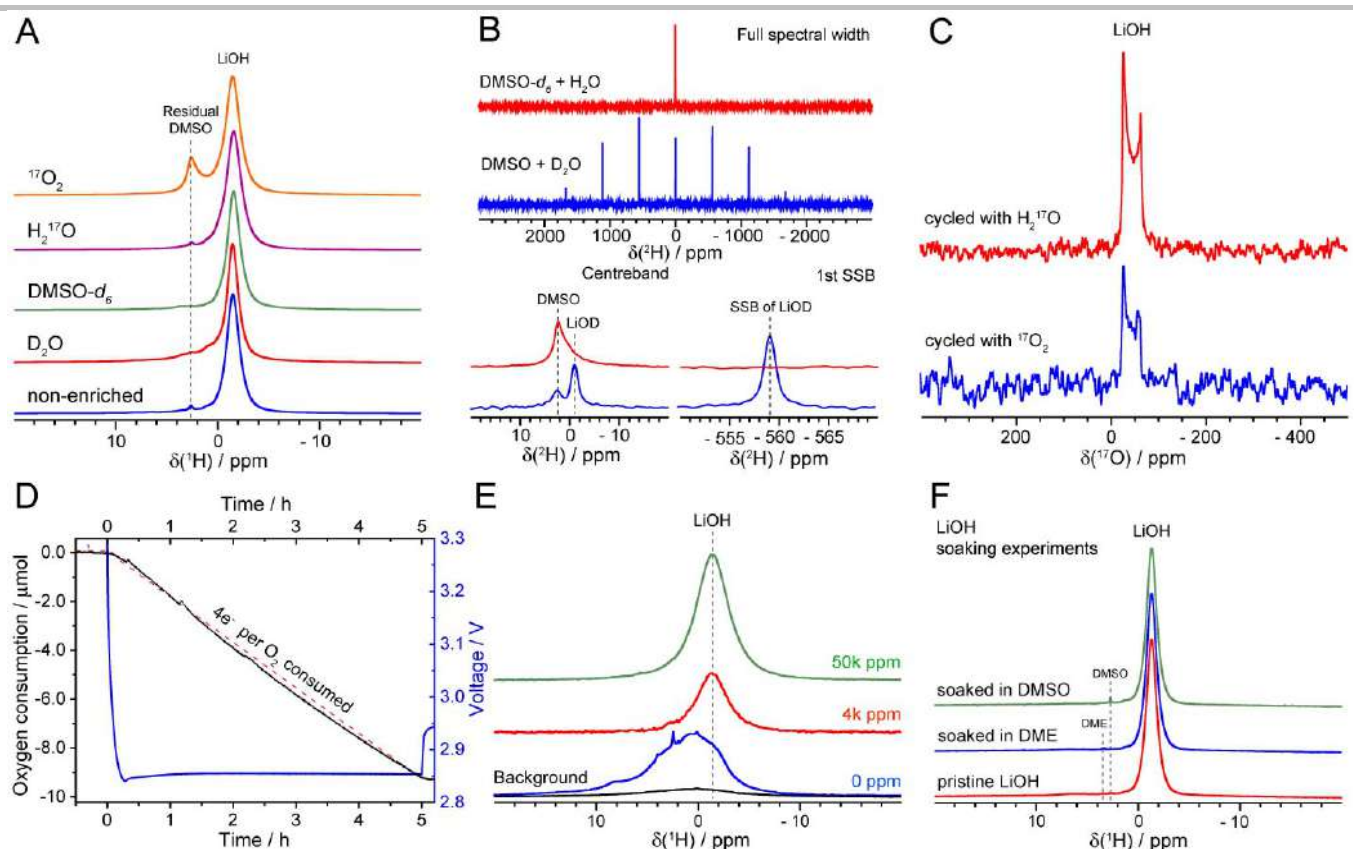

Fig. 2 [1]H (A), [2]H (B) and [17]O (C) solid state NMR spectra of discharged Ru/SP electrodes, prepared from Li-O₂ cells with 1 M LiTFSI/DMSO electrolyte with 4000 ppm water. The [1]H NMR spectra (A) show that all samples, independent of the nature of isotope enrichment (as labelled), give rise to a dominant resonance at -1.5 ppm corresponding to LiOH; the small resonance at 2.5 ppm is due to residual DMSO. The [2]H NMR spectra confirm that water is the proton source for LiOH formation. Note that the [1]H NMR experiments in Fig. 2A are not quantitative, and the LiOH detected in the case with added D₂O in the DMSO-based electrolyte is likely due to H₂O impurities from D₂O. *Operando* pressure measurement of a Ru-catalyzed cell with 50,000 ppm water (D) and quantitative [1]H NMR spectra (E) of 1st discharged electrodes prepared from Li-O₂ cells using 1 M LiTFSI/DMSO electrolyte with 0, 4000, and 50000 ppm water contents. 10 μmol O₂ consumption corresponds to 27.7 mbar pressure drop measured for 1 mAh capacity (200 μA, 5 hours). [1]H NMR evaluating the long-term stability of LiOH in DMSO and DME solvents (F) by comparing LiOH powder with those after being soaked in DMSO and DME solvents for a month. Apart from the residual DMSO or DME solvent, no additional signals are observed, the soaked LiOH powder remaining chemically unchanged.

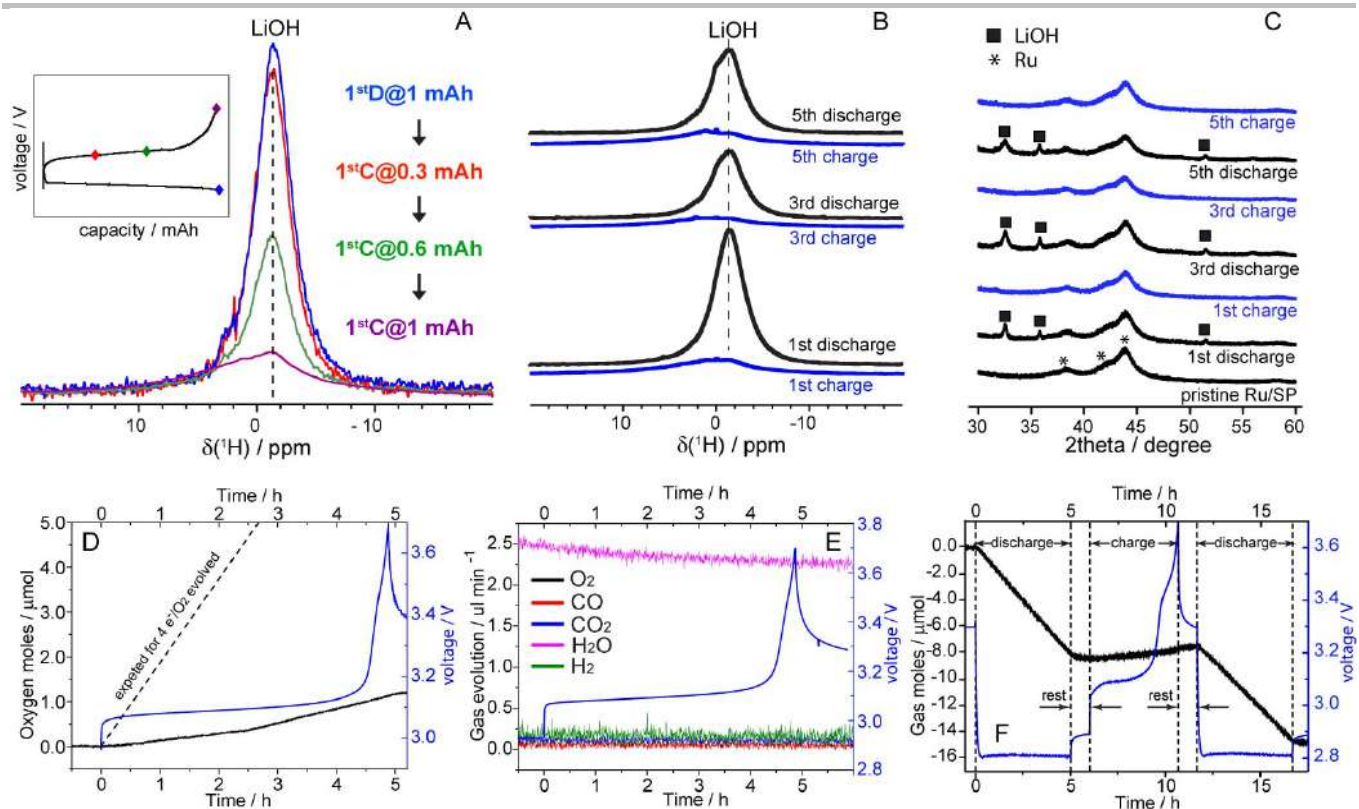

Fig. 3 Quantitative ¹H (A, B), *ex situ* XRD measurements (C) of cycled Ru/SP electrodes prepared using 1 M LiTFSI/DMSO electrolytes with 50,000 ppm water; *operando* electrochemical pressure (D, F) and mass spectrometry (E) measurements of a Ru-catalyzed cell with 50,000 (D) and 4000 ppm (F) water. Batteries terminated both at different state of charge (A) and fully charged following different discharge-charge cycles all show quantitative electrochemical removal of LiOH. Little $O_2$ evolution is seen during charging (D, E) and the cell pressure continues to drop over extended cycles (F). 10 μmol $O_2$ in D and F corresponds to 27.7 mbar pressure change measured for 1 mAh capacity (200 μA, 5 hours).

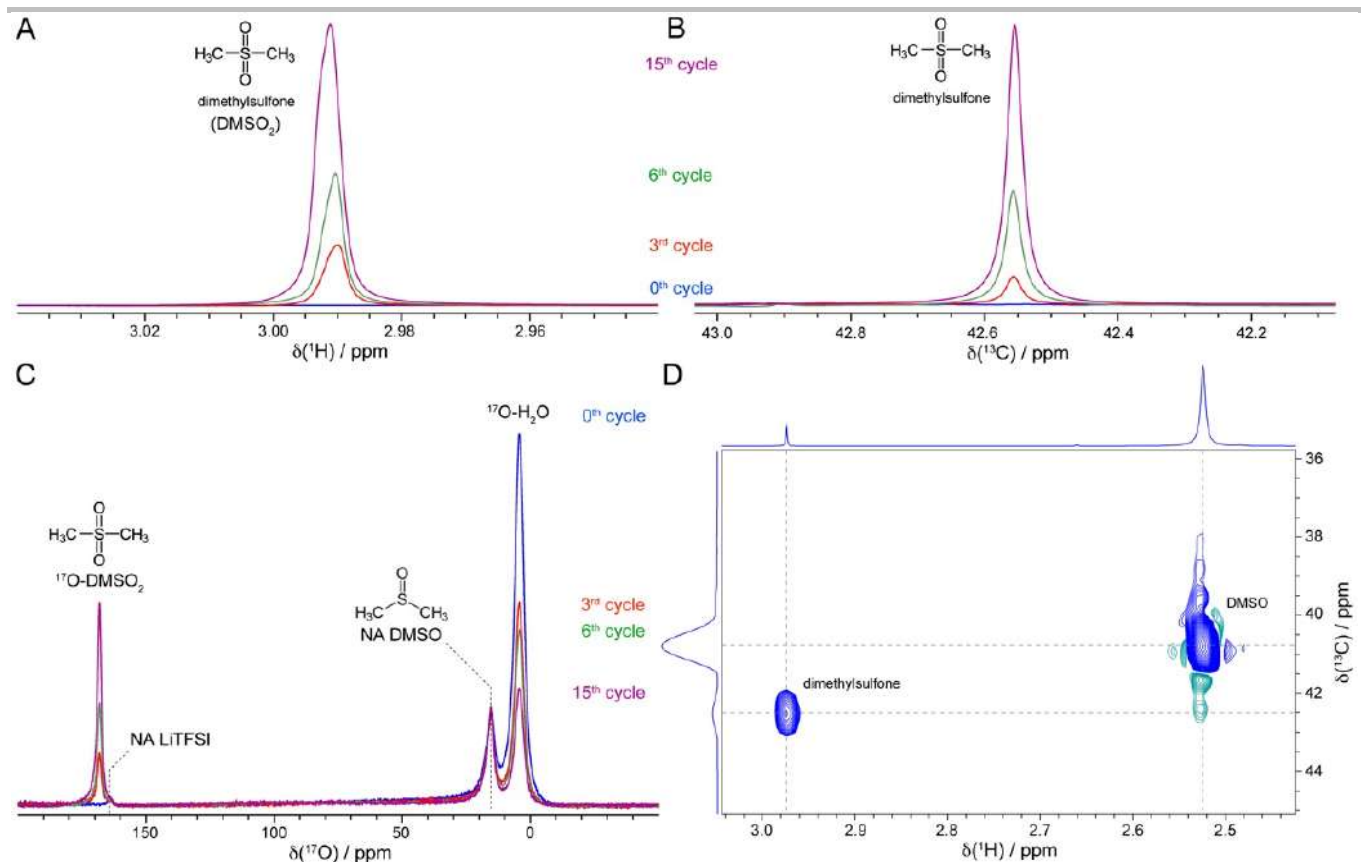

Fig. 4 $^1H$ (A), $^{13}C$ (B) and $^{17}O$ (C) and $^1H$-$^{13}C$ heteronuclear single quantum correlation (D) solution NMR spectra of cycled 1M LiTFSI/DMSO electrolyte with 45,000 ppm $^{17}O$ enriched water from Ru-catalyzed Li-O$_2$ batteries. New resonances at 2.99 ppm ($^1H$), 42.55 ppm ($^{13}C$) and 169 ppm ($^{17}O$) signify the formation of DMSO$_2$. The heteronuclear correlation experiment was performed on a charged electrolyte at the end of the 6$^{th}$ cycle. The cross peak at (2.99 ppm $^1H$ – 42.55 ppm $^{13}C$) further supports DMSO$_2$ formation; the other cross peak at (2.54 ppm $^1H$ - 41.0 ppm $^{13}C$) is due to DMSO.